\documentstyle{lamuphys}

\begin{document}
\title{ Supergeometry in Equivariant Cohomology}
\author{Armen Nersessian\inst{1}\inst{2}}
\institute{ Bogolyubov Laboratory of Theoretical Physics, JINR,\\
Dubna, 141980, Russia
\and
 Department of Theoretical Physics, Yerevan State University, \\
A. Manoukian st., 5, Yerevan, 375012 Armenia\\
E-mail: {\tt nerses@thsun1.jinr.ru}}
\maketitle
\begin{abstract}
We analyze  $S^1$ equiva\-riant cohomology from the su\-per\-geomet\-rical
point of view. For this purpose we equip the
 external algebra  of given manifold with
equivariant even su\-per\-(pre)symp\-lec\-tic struc\-tu\-re, and
 show, that its Poincare-Car\-tan inva\-riant defines  equi\-va\-ri\-ant
Euler classes of  surfaces.
This allows to derive localiza\-tion formulae by  use
of superanalog of Stockes theorem.
\end{abstract}
 \setcounter{equation}0
\section{Introduction}
Since the late  eighties localization formulae attract permanent interest
in the physical community due their application to evaluation
of path integrals in quantum mechanics,
 topological and supersymmetric field theories
(see \cite{niemi} and refs therein).

The original localization formula \cite{DH} states,
that if $(M,\omega, S^1)$ is  $2n$-dimensional compact symplectic manifold,
and $H(x)$ is the Hamiltonian, generating  $S^1$-group symplectic
action, the classical partition function  is localized over
 the critical points  of Hamiltonaian
 \begin{equation}
\left(\frac{i\phi}{{2\pi}}\right)^{2n}
\int_M{\rm e}^{i\phi H}\frac{\omega^n}{n!}=
\sum_{dH=0}\frac{{\rm e}^{i\phi H}
{{\sqrt{\det\omega}}}}{\sqrt{\det {\rm Pf} H}}.
\label{DH}\end{equation}
Application  of this formula to path integrals gives an elegant way
for their evaluation,
 as well as the conditions of exactness of stationary phase approximation.

Localization formula of Duistermaat and Heckman
 can be naturally interpreted in terms of
equivariant cohomology
\footnote{$G$-equivariant cohomology of the $G$-manifold $(M,G)$
called the $G$-invariant cohomology of the factormanifold $M/G$ }
\cite{atiah}.
This   allowed one to obtain other localization formulae,
related with both Abelian (see, e.g. \cite{berline}),
 and non-Abelian \cite{witten} equivariant cohomologies.
Also other features of equivariant cohomologies
 relevant to various  aspects of quantum
field theory were found (see \cite{niemi,witten,aj} and refs therein).

The language of supersymmetry  simplifies  greatly
formulation  of  equivariant cohomologies,
openings  new horizons in their study by use of
advanced supermatematical technique
\cite{zs}. It gives the shortest way for incorporation of
equivariant cohomology in supersymmetric theories and
BRST quantization methods.

In the framework of supermathematical  description of equivariant
cohomology   the exterior  algebra $\Lambda M$
 of the given manifold $M$  is considered  a supermanifold,
 parametrized   by local coordinates $z^A =(x^i, \theta^i)$,
 where $x^i$ are local coordinates of $M$ and
$\theta^i$ are basic 1-forms $dx^i$, $p(\theta^i)= 1$.
Thus,  differential forms on $M$ can be considered  as
 functions defined on a supermanifold.
The  operators of  exterior derivative, Lie derivative and
inner product are represented by vector fields.
Divergency operator is represented by
 odd second-order differential operator, known as
``Batalin-Vilkovisky $\Delta$-operator".

 Most  convenient model of $S^1-$equivariant cohomology,
 the so-called Cartan model,
is formulated  in terms of
simplest superalgebra
\begin{equation}
 [\hat E ,\hat E ]_{+} =2\hat X,
\label{alg}\end{equation}
realized by the vector fields:
\begin{equation}
\label{E}{\hat X}=\xi^i(x)\frac{\partial}{\partial x^i} +
\xi_{,k}^{i}(x)\theta^k
\frac{\partial}{\partial \theta^i},\quad {\hat E} =\xi^i(x) \frac{\partial}{%
\partial \theta^i} +\theta^i \frac{\partial}{\partial x^i}
\end{equation}
where vector field $\xi=\xi^i\frac{\partial}{\partial x^i}$
defines infinitesimal  $S^1$-action  on $M$.

It is obvious that vector $\hat X$ corresponds to
Lie derivative of differential forms. The vector $\hat E$
corresponds to sum of  operators of exterior derivative  and inner product.
Expression  (\ref{alg}) is nothing else  but the  homotopy
formula $ L_{\xi}= d\imath_{\xi} +\imath_{\xi}d$.
Thus, restriction of   $\hat E$ field (equivariant differential)
 to the (sub)space of ${\hat X}$-invariant functions is nilpotent, and
its cohomologies on this subspace defines
the equivariant cohomologies of $(M,S^1)$.

To relate equivariant cohomologies with localization formulae, let
consider the following functional
\begin{equation} \label{invint0} Z^{\lambda}(A)=
\int_{\Lambda(M)}A(x,\theta){\rm e}^{-\lambda{\hat E}\Psi}{\cal  D}(x,\theta),
\end{equation}
where ${\cal D}$ is  volume element, $A$ and  $\Psi$ are respectively
even and odd functions on
$\Lambda M$, $\lambda$ is arbitrary parameter.

This functional  is ${\hat E}$-invariant, if  requires:
\begin{equation}
  {\hat E} A =0,
\quad{\hat X}\Psi=0,\quad div_{\cal D} {\hat E}=0.
\end{equation}
Applying standard BRST analyzes, one can find, that the functional
(\ref{invint0})  is  $\lambda$-independent, i.e. only
$S^1$-equivariant forms  have contribution  to its value.

In order to derive localization formulae, we
have to choose the "gauge fermion"
\begin{equation}
 \Psi=\xi^ig_{ij}\theta^i,
\label{psi}\end{equation}
where $g$ is an  $S^1$-invariant Riemann metric: ${\cal L}_\xi g=0$.

Taking into account the following representation
 of $\delta$-function
\begin{equation}
\delta(\xi)=\frac{\lambda^{{\rm dim}\,M}}
{\pi^{{\rm dim}\,M}\sqrt{\det g_{ij}}}
\lim_{\lambda\to\infty}{\rm e}^{-\lambda\xi^i g_{ij}\xi^j}
\end{equation}
we find, that  in  $\lambda\to\infty$ limit the functional
(\ref{invint0}) localizes in the critical points of $\xi$.

For example, for a compact symplectic manifold $(M, \omega, S^1)$,
where $\xi=\omega^{-1}(dH,\;\;)$ one can
 choose ${\cal D}=1$ and $A$ is equivariant Chern class
$Ch(H,\omega)$  defined as follows
\begin{equation}
Ch(H,\omega)\equiv\exp i\phi (H-\frac 12\omega_{ij}\theta^i\theta^j).
\label{chern}\end{equation}
 Substituting these expressions in (\ref{invint0}), we will get
Duistermaat-Heckman formula (\ref{DH}).

There is also another ${\hat E}$-invariant function, $Eu(\xi,g)$
 \begin{equation}
{Eu}(\xi, g)\equiv
{\sqrt{\det (\xi^i_{;j}+ R^{i}_{jkl}\theta^{k}\theta^{l})}},
\label{eu}\end{equation}
called equivariant Euler class.

Substituting this function in (\ref{invint0}), one gets that in
$\lambda\to 0$ limit  we recognize the statement of
 Poincare-Hopf theorem ,
while in $\lambda\to\infty$ limit: Gauss-Bonnet Theorem.\\
Derivation of another existing localization formulae is analogous.

In present paper we  study  $S^1$ equivariant cohomology
from the point of view of supersymplectic geometry.
The key observation is that on $\Lambda M$ there exists
 ${\hat E}$-invariant presymplectic structure,
whose   Poincare-Cartans ivariants
define equivariant Euler classes of the surfaces in given manifold.
This allows one to formulate an analog of the functional
(\ref{invint0}) on surfaces in $\Lambda M$,
and to  derive localization formulae
by use of generalization of  Stokes theorem
for  even symplectic supermanifolds \cite{kst}
(Khudaverdian, Schwarz and Tyupkin),
and relate them with topological numbers of surfaces.
In a contrast to equivariant Euler classes, equivariant
Chern classes  are  superfunctions, related with the
odd symplectic structure, constructed on the
exterior algebra of symplectic manifold \cite{jetp}.

Remarkable point of actual approach is that  it is based on
the odd symplectic geometry, which gives transparent
parallels with  Batalin-Vilkovisky formalism \cite{bv}.

This make possible the application to equivariant
cohomologies the results, concerning the geometry
of Batalin-Vilkovisky formalism (\cite{schwarz}).
From the other hand, using of  recently developed
technique of quantization of antibrackets \cite{bm}
seems  fruitful  in application of equivariant cohomology
to path integrals.
\section{Equivariant Euler Class.}
Let us consider $(M, S^1)$ manifold with the  $S^1$-invariant metric  $g$.
Using this metrics, let us construct  on   $\Lambda M$ an
 odd ${\hat X}$-invariant one-form and
corresponding  odd symplectic structure:
\begin{eqnarray}
&{\cal A}=\theta^ig_{ij}dx^j ,& {\cal L}_{\hat X}{\cal A}=0;\\
& \Omega_1=d{\cal A}=g_{ij}dx^{i}\wedge D\theta^{j},
&{\cal L}_{\hat X}\Omega_1=0,\label{ogs}\end{eqnarray}
where
\begin{equation}
D\theta^{j}\equiv
g_{ij}dx^{i}\wedge (d\theta  ^i+\Gamma _{kl}^i\theta ^kdx^l),
\end{equation}
and   $\Gamma_{kl}^i$ denote Cristoffel symbols of the metrics $g$.

From this follows, that   ${\hat X}$ is hamiltonian
vector field one with respect to (\ref{ogs}),
and the gauge fermion (\ref{psi}), plays the role of Hamiltonian:
\begin{equation}
d\Psi=\Omega_1(\hat X,\quad),\quad\Psi=\xi^ig_{ij}\theta^j .
\end{equation}

Forms ${\cal A}$ and $\Omega_1$ possess remarkable property:
being ${\hat X}-$ invariant, they are not  ${\hat E}$-invariant:
\begin{equation}
{\cal E}\equiv{\cal L}_{\hat E}{\cal A}\neq 0;
\quad  \Omega_0\equiv {\cal L}_{\hat E}\Omega_1 \neq 0.
\end{equation}
Explicitly,
 \begin{eqnarray}
&{\cal E}=\xi_i dx^i +\theta^i g_{ij}D\theta^j ,&\label{calE} \\
&\Omega_0 =
 \frac 12 (\xi_{[i,j]}+g_{in}R^n_{jkl}\theta ^k\theta^l)dx^i\wedge dx^j
+g_{ij} D\theta^i\wedge D\theta^j .&   \label{evens}
\end{eqnarray}
From Eqs. (\ref{calE}) and (\ref{evens})  it follows immediately,
that the forms ${\cal E}$ and ${\cal\Omega}$ are
 ${\hat E}$-invariant.
Indeed
$$ {\cal L}_{\hat E }{\cal E}={\cal L}_{\hat E}
{\cal L}_{\hat E}{\cal A}=2{\cal L}_{\hat X}{\cal A}=0. $$
Similarly one can check ${\hat E}$-invariance    of the second structure.

Since the  two-form ${\cal \Omega}$  is closed, we conclude,
that {\it the expression (\ref{evens}) defines
the  $S^1$-equivariant even pre-symplectic structure
on $\Lambda M$.}

The  Hamiltonians of the vector fields, given by Eq.(\ref{E}) read:
\begin{eqnarray}
&\Omega_0(\hat X,\quad)=d{\cal H}, \quad
{\cal H}={\hat E}\Psi=\xi^i g_{ij}\xi^{j}-\xi_{i;j}\theta^i\theta^j.&\\
&\Omega_0(\hat E,\quad)=d\Psi,\quad \Psi=\xi_i\theta^i .&
\end{eqnarray}
Consider now a closed surface $\Gamma\subset\Lambda M$,
$\Omega\vert_\Gamma\neq 0$,
parametrized by the equations $z^{A}=z^{A}(w)$,
where $w^\mu$ are local coordinates of $\Gamma$.

One can construct   a closed density on $\Gamma$,
invariant under canonical transformations of $\Omega_0$
(Poincare-Cartan invariant), which defines characteristic class of $\Gamma$
\cite{kst}:
  \begin{equation}\label{bst}
{\cal D}_{\Gamma}(z(w),\partial z(w))=
\sqrt{Ber \frac{\partial z^A}{\partial w^\mu}\Omega_{(0){AB}}
\frac{\partial z^B}{\partial w^\nu}} .
\end{equation}

Taking into account, that  ${\hat E}$-field generates canonical
transformation, we conclude, that the density,
given by Eq.(\ref{bst}) is  ${\hat E}-$invariant.
Thus, it defines an   equivariant characteristic class of $\Gamma$ surface.
When $\Gamma$ coincides with $\Lambda M$, this density
is just {\it equivariant Euler class}  (\ref{eu}) .

Let us give some explicit realization of  density, given by
Eq. (\ref{bst}).
Consider  surface $\Gamma\subset\Lambda M$, given  by
the equations
\begin{equation}
z^{A}=z^{A}(w):\quad x^i=x^i (y^a), \quad
\theta^i = P^i_\alpha(y)\eta^\alpha, \quad\quad  p(\eta)=p(y)+1.
\end{equation}
Here $w^\mu =(y^a,\eta^\alpha )$ are local coordinates on
$\Gamma$,  where $y^a$ are  local coordinates on $N\subset M$
(notice, that after change of the grading
$ p(\eta)\to p(\eta)+1$ lead to vector bundle $V(N)\subset T(M)$,
associated  with $\Gamma$).

The restriction of presymplectic structure (\ref{evens}), on  $\Gamma_0$
looks as follows
\begin{equation}
\Omega_0\vert_{\Gamma_0} = \frac 12 (\xi_{[a,b]}+
g_{\alpha\delta}R^{\delta}_{\beta ab}\eta^{\alpha}\eta^{\beta} )dy^a\wedge
dy^b + g_{\alpha\beta}D\eta^{\alpha}\wedge D\eta^{\beta},
\end{equation}
where
$$
D\eta^\alpha\equiv d\eta^\alpha + A^\alpha_{\beta}\eta^\beta,\quad
A^\alpha_{\beta}=g^{\alpha\delta}P^i_\delta g_{ij}\left
(P^j_{\beta,a}+\Gamma^j_{lk}P^k_\beta \frac{\partial x^l}{\partial y^a}
\right)dy^a, $$
  defines connection one-form,
compatible with  the induced metric
on fiber, given by the following expression
$$g_{\alpha\beta}=P^i_\alpha g_{ij}P^j_\beta,$$
 $R^{\delta}_{\beta ab}dy^a\wedge dy^b$ is the curvature of
 the connection  $A^\alpha_{\beta}$,
while
$$
\xi_{[a,b]}dy^a\wedge dy^b\equiv
\xi_{i;j}\frac{\partial x^i}{\partial y^a}\frac{\partial x^j}{\partial y^b}
dy^a\wedge dy^b $$
is  induced (pre)symplectic structure on $N$.

Thus, on the surface $\Gamma_0$ the density (\ref{bst})
takes the following form
\begin{equation}
 \label{d} {\cal D}(\Omega |\Gamma_0, w)=
 \left(\frac{\det ( \xi_{[a,b]}+
g_{\alpha\delta}R^{\delta}_{\beta ab}\eta^{\alpha}\eta^{\beta})}{\det
g_{\alpha\beta}}\right)^\frac{1}{2}.\end{equation}

\section{Equivariant Chern Class.}
The supergeometrical origin of equivariant Chern class,
given by Eq.(\ref{chern}) considerably  simpler, than Euler one.
Chern class is a  function  on the  exterior algebra of
symplectic manifold $(M,\omega,S^1)$, with  $S^1-$group action,
 generated  by the Hamiltonian $H$: $\xi=\omega^{-1}(dH, \;)$.

This algebra can be equipped with odd ${\hat E}$-invariant symplectic
structure $\Omega$, in contrast to $\Omega_1$, given by Eq.(\ref{ogs})
\cite{jetp}. Indeed,
$${\cal L}_{\hat X}\omega=0,\quad
{\cal L}_{\hat E}\omega\equiv\Omega\neq 0,\;\;\Rightarrow
{\cal L}_{\hat E}\Omega=\frac 12 {\cal L}_{\hat X}\omega=0,
\quad d\Omega=0.$$
Explicitly, $\Omega$ looks as follows
\begin{equation}
\Omega=\omega_{ij}dx^i\wedge d\theta^j +\frac 12 \omega_{ij,k}\theta^k
dx^i\wedge dx^j .\label{myb}\end{equation}
The Hamiltonians of the vector fields ${\hat E}$ and ${\hat X}$
are defined  by the following expressions:
\begin{equation}
d(H-F)=\Omega({\hat E, \;\;}),\quad Q=\Omega({\hat X},\;\;),
\end{equation}
where
$$F=\frac 12 \omega_{ij}\theta^i\theta^j, \;\;Q={\hat E}H=
\theta^i{\partial H}/{\partial x^i}.$$
Functions $F,\; Q,\; H$, form superalgebra
\begin{equation}
\begin{array}{c}
\{H\pm F, H \pm F\} =\pm 2Q , \quad\{H + F , H - F\} =0\\
  \{H\pm F, Q \} =
\{ Q, Q\} = 0,
\end{array}\label{susy1}\end{equation}
where    $\{\;\;,\;\;\}$ denotes the antibracket, corresponding to the
odd symplectic structure $\Omega$ given by Eq. (\ref{myb}).

The  functions $H,F,Q$ forms  algebra (\ref{susy1}) with respect to
odd symplectic structure given by Eq.(\ref{ogs}), only if
$(M,g,\omega)$ is a Kahler manifold.
This is  the only case, when  antibrackets,
corresponding to these symplectic structures are compatible in the sense
of bi-Hamiltonian mechanics.
Other relations of equivariant Chern classes with supersymmetric
mechanics can be found in Ref.\cite{asi}.
\section{Localization Formulae.}
 Above we constructed ${\hat E}$-inva\-riant closed density
(\ref{bst}), cor\-res\-pon\-ding to equi\-va\-riant Euler classes.
Using this density, we can
define the following functional
\begin{equation} \label{dinvint2}
Z^{\lambda}(\Gamma, A)=
\int_{\Gamma}A(z){\rm e}^{-\lambda{\hat E}\Psi} {\cal D}_{\Gamma}
\end{equation}
on  the surface $\Gamma$.\\
This functional become ${\hat E}$-invariant,
if we assume, as in Introduction,
${\hat E}A={\hat X}\Psi=0$.
From ${\hat E}$-invariance  follows its $\lambda$-independence.

Since the density (\ref{bst}) is closed, it doesn't
change under smooth deformation of $\Gamma$ if  $A=1$ \cite{kst}, i.e.
it is defines  a topological invariant.
In  $\lambda\to 0$ and $\lambda\to\infty$ limits
one reproduces the statements of Gauss-Bonnet and
Poincare-Hopf theorems for  Euler characters of the surfaces.
Choosing $A=Ch(H,\omega)$  one obtains a generalization of
these statements, while $A=Ch(H,\omega)Eu^{-1}(M,g)$ corresponds to
 Duistermaat-Heckman  formula for a surface.

Using the functional (\ref{dinvint2}), one can easily derive localization
formulae for the case, when the critical points of $\xi$-field
form some submanifold  $M_0$ \cite{niemi2}(Niemi and Palo, 1994).
 Consider the decomposition $M=M_0\cup N$, where
 $M_0$ is the set of critical points of $\xi$ and
$N$ is normal bundle to $M_0$.
Under  appropriate choice of  $A(z)$ discussed above,
one comes to formulae, derived in \cite{niemi2}.

Finally, let us give the dual construction for the functional
(\ref{dinvint2}), when  critical points of  $\xi$ are isolated.
Let   $\Gamma$ be parametrized by the set equations
$f^a (z)=0$, $a=1,...{\rm codim}\;\Gamma $.
In this case the functional   (\ref{dinvint2})
can be presented in the following form
\begin{equation}
\label{dinvint}
Z=\int_{\Lambda M}A(z)\delta(f^a)
\sqrt{{\rm Ber}\{f^a, f^b\}_0} {\cal D}_{\Lambda M}(z),\end{equation}
 where $\{f(z),g(z)\}_0$ is  Poisson bracket,
cor\-res\-pon\-ding to the (pre)symp\-lec\-tic 2-form (\ref{evens}). \\

{\Large{Acknowledgments}}\\
 The author would like to thank I.A.Batalin, O. M. Khudaverdian, A. Niemi
 for the interest to  work and useful comments.

Special thanks to C.Sochichiu for careful reading of
 manuscript and numerous  remarks.

This work has been supported in part by grants
 INTAS-RFBR No.95-0829, INTAS-96-538 and INTAS-93-127-ext .

\end{document}